# Strain-induced anion ordering in perovskite oxyfluoride films

Jiayi Wang[1], Yongjin Shin[2], Jay R. Paudel[3], Joseph D. Grassi[3], Raj K. Sah[3], Weibing Yang[3], Evguenia Karapetrova[4], Abdulhadi Zaidan[5], Vladimir N. Strocov[5], Christoph Klewe[6], Padraic Shafer[6], Alexander X. Gray[3,7], James M. Rondinelli[2], Steven J. May[1,*]

[1] Department of Materials Science and Engineering, Drexel University, Philadelphia, Pennsylvania 19104, USA

[2] Department of Materials Science and Engineering, Northwestern University, Evanston, Illinois 60208, USA

[3] Department of Physics, Temple University, Philadelphia, Pennsylvania 19122, USA

[4] Advanced Photon Source, Argonne National Laboratory, Argonne, Illinois 60439, USA

[5] Swiss Light Source, Paul Scherrer Institute, CH-5232 Villigen, Switzerland

[6] Advanced Light Source, Lawrence Berkeley National Laboratory, Berkeley, California 94720, USA

[7] Temple Materials Institute, Temple University, Philadelphia, Pennsylvania 19122, USA

**ABSTRACT**

Anionic ordering is a promising route to engineer physical properties in functional heteroanionic materials. A central challenge in the study of anion-ordered compounds lies in developing robust synthetic strategies to control anion occupation and in understanding the resultant implications for electronic structure. Here, we show that epitaxial strain induces preferential occupation of F and O on the anion sites in perovskite oxyfluoride SrMnO$_{2.5-\delta}$F$_\gamma$ films grown on different substrates. Under compressive strain, F tends to take the apical-like sites, which was revealed by F and O $K$-edge linearly polarized x-ray absorption spectroscopy and density functional theory calculations, resulting in an enhanced $c$-axis expansion. Under tensile strain, F tends to take the equatorial-like sites, enabling the longer Mn-F bonds to lie within the plane. The anion ordered oxyfluoride films exhibit a significant orbital polarization of the 3*d* electrons, distinct F-site dependence to their valence band density of states, and an enhanced resistivity when F occupies the apical-like anion site compared to the equatorial-like site. By demonstrating a general strategy for inducing anion-site order in oxyfluoride perovskites, this work lays the foundation for future materials design and synthesis efforts that leverage this greater degree of atomic control to realize new polar or quasi-two-dimensional materials.

1. **INTRODUCTION**

Inducing atomic-scale ordering is a promising approach for realizing new or enhanced physical properties in functional materials with mixed cation and anion site occupation. Perovskites, which exhibit the general chemical form $ABX_3$, have been recognized as a versatile family of functional materials due to their structure that accommodates a wide range of chemistries: the *A*-site cations are coordinated with 12 *X*-site anions; the *B*-site cations are coordinated with 6 *X*-site anions forming an $BX_6$ octahedral structure, which shares corners to form a three-dimensional network.[1,2] In this family, cation substitution is a well-established route to engineer the properties of perovskites as the *A*- or *B*-sites can be readily occupied by two or more different elements.[3] Beyond solid solutions, long-range ordering of cations can enhance or endow new functionality into quaternary perovskites by removing disorder or enforcing polarity.[4-6] Alloying on the anion-site provides an alternative means of modulating the properties of perovskite oxides, as heterovalent anionic substitution not only alters the electronic state of *B*-site cation, but also the *B-X* bonding geometry and ionicity due to the different anionic characteristics, such as electronegativity, polarizability, charge number and atomic size. Previous work has demonstrated that a wide range of oxynitrides, oxyhydrides, and oxyfluorides can be synthesized, in which $N^{3-}$, $H^-$, and $F^-$ anions are incorporated into the lattice due to their similar radii as $O^{2-}$.[7-10] However, compared to approaches based on cationic substitutions, heteroanionic-based materials design is at a less mature state and fundamental questions remain about how to deterministically control the concentration and site-occupancy of the different anionic species.[11]

Epitaxial thin films provide new avenues to enable synthetic control of heteroanionic materials, namely through the formation of interfaces and substrate-induced biaxial strain. For example, biaxial strains of ±3% are now routinely imposed in epitaxial oxide films that are tens of nm in thickness without loss of coherence with the substrate.[12] The development of topochemical approaches[13-15] to convert perovskite oxide films to oxyfluorides has enabled the realization of numerous families of epitaxial oxyfluorides including ferrites,[14-18] manganites,[19,20] ruthenates,[21,22] chromates,[23] cobaltites,[24] and nickelates.[25] The thin film nature of these samples has been exploited for functionality not accessible to bulk polycrystalline oxyfluorides, for instance through reversible insertion and removal of F in $NdNi(O,F)_{3-\delta}$ films[25] and lateral patterning of $SrFeO_{2.5}/SrFeO_2F$ heterostructures through spatially controlled topochemistry.[26] However, the use of substrate-induced strain to alter or direct the F incorporation has yet to be reported.

Here, we demonstrate that epitaxial strain can be used to control F/O ordering in epitaxial $SrMnO_{2.5-\delta}F_\gamma$ (SMOF) thin films deposited on substrates with different in-plane lattice constants. By altering the in-plane and out-of-plane *B-X* bond lengths, this strain has considerable consequences for the relative formation energies of *B*-F and *B*-O bonds and thus the site occupancy of the F and O anions. Using a complementary suite of x-ray linear dichroism, x-ray diffraction, polarization-dependent resonant valence-band photoemission, hard x-ray photoemission, and density functional theory calculations, we demonstrate that F occupies the apical-like sites (O1-sites) under compressive strain leading to Mn-F bonds oriented along the out-of-plane direction. In contrast, F is incorporated on equatorial-like sites (O2, O3-sites) under tensile strain, leading to Mn-F bonds parallel to the in-plane directions. An extreme x-ray linear dichroism at the F and O $K$-edges results from the anionic configurations within the films. This work highlights a unique opportunity for materials design utilizing topotactic modification of epitaxial films, which enables static strain states that are inaccessible in bulk geometries thus providing a route to metastable ordered structures.

## 2. RESULTS AND DISCUSSION

The anisotropic deformation of as-grown $SrMnO_{2.5}$ (SMO) films upon biaxial strain is shown in Figure 1(a,b). Bulk SMO is orthorhombic, but can be approximated as having a pseudocubic lattice parameter of 3.811 Å[27,28], which results in a biaxial compressive strain when grown on $LaAlO_3$ (LAO; $a = 3.795$ Å), and a tensile strain when deposited on $(La,Sr)(Al,Ta)O_3$ (LSAT; $a = 3.870$ Å) and $SrTiO_3$ (STO; $a = 3.905$ Å) substrates. For SMO/LAO, the in-plane lattice parameter is compressed, which induces a lattice elongation along $c$-direction. For the SMO/LSAT and SMO/STO, the in-plane lattice is elongated, and the $c$-axis parameter is contracted. Figure 1(a) also shows the positions of the O1, O2, and O3 sites. The O1 site forms Mn-O-Mn bonds along the out-of-plane direction (apical-like), while the O2 and O3 sites participate in in-plane Mn-O-Mn bonds (equatorial-like).

Figure 1(c) shows the $c$-axis parameters of as-grown SMO and fluorinated SMOF on LAO, LSAT, and STO substrates, obtained from diffraction data presented in supplemental Table S1 and Figure S1. The $c$-axis parameter expands when the film is compressed on LAO ($a_{bulk} > a_{LAO}$) and contracts for the tensile strained SMO grown on LSAT and STO. After fluorination, there is a $c$-axis expansion for all films, consistent with other reports of lattice expansion upon conversion of oxides to oxyfluorides in which the F is incorporated into the perovskite lattice.[15,19,29] Reciprocal space maps, shown in Figure S1, confirm that the SMOF films remain strained to the substrate following the topochemical fluorination reaction. Comparing all three fluorinated films, SMOF/LAO shows a much larger $c$-axis expansion than SMOF on LSAT and STO.

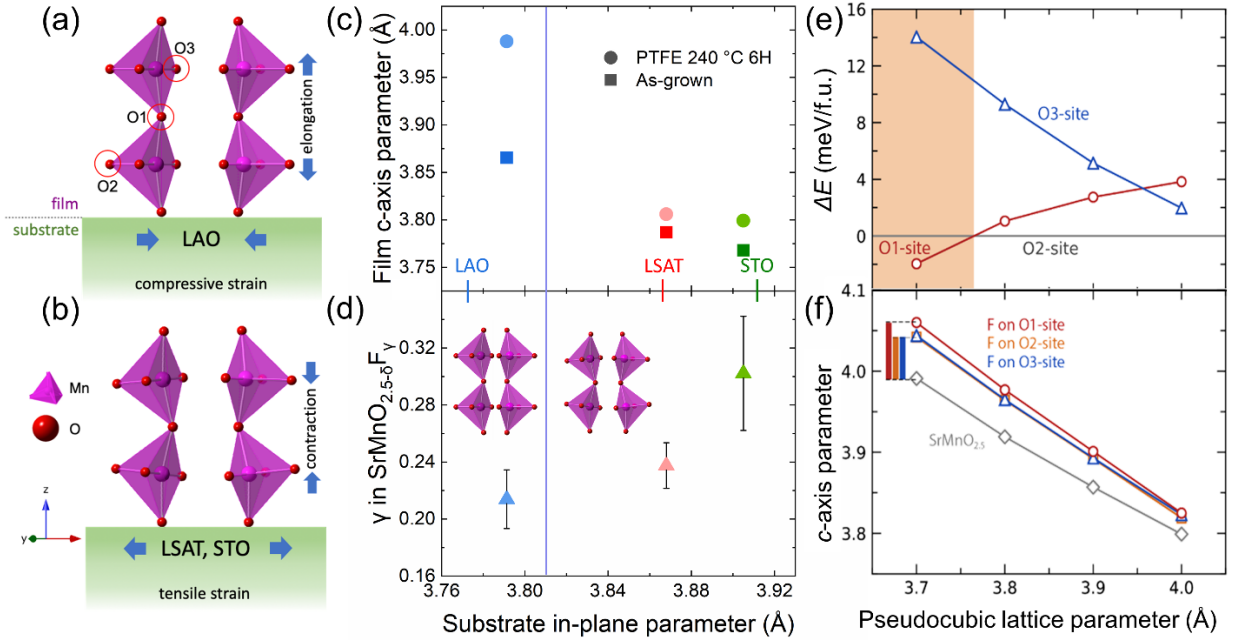

**Figure 1.** Impact of strain on the structural properties, composition, and F-site occupancy. Schematics illustrating the anisotropic deformation of $SrMnO_{2.5}$ under (a) compressive and (b) tensile strain. (c) The $c$-axis parameters of the films before and after fluorination. (d) F content ($\gamma$) of the $SrMnO_{2.5-\delta}F_\gamma$ films grown on LAO, LSAT, and STO. Error bars indicate the standard deviation of the average F concentration from the XPS sputter depth profile. Data represented in the same color are collected from the same sample before and after fluorination. The calculated (e) relative energies and (f) $c$-axis parameter of different F occupancies as a function of the in-plane pseudocubic lattice parameter. In panel (e), the energy of the O2-site substitution is set as the reference state.

The impact of strain on F incorporation was measured using x-ray photoemission spectroscopy (XPS) to quantify the F concentration within the samples and to confirm that the F extents through the films. In Figure 1(d), the F content of the SMOF films is plotted as a function of the substrate in-plane parameter. The compressively strained SMOF/LAO shows the lowest F concentration, indicating that the large $c$-axis enhancement in this sample upon fluorination is not the result of an increased F incorporation compared to the films under tensile strain. We attribute the differences in F content to strain-induced differences in the F$^-$ diffusion rate along the pseudocubic [001] direction, as well as the lattice expansion that occurs upon F$^-$ incorporation. Given that the F incorporation process proceeds from the surface vertically through the $c$-axis oriented films, we hypothesize that the smaller in-plane lattice parameters of SMOF/LAO impede vertical F diffusion by increasing the atomic density within the (001) planes, while simultaneously the necessary lattice expansion upon F$^-$ incorporation is unfavorable due to the compressive strain imposed by the substrate. Under increasing tensile strain, the energetic and diffusion considerations are opposite to those described above due to the larger in-plane lattice constants, and thus the greater F content inside the films on LSAT and STO. In this way, the lattice expanding effects of F$^-$ incorporation results in trends analogous to oxygen vacancy formation in strained perovskites, which is promoted under tensile strain but not under compressive strain.[30,31]

Density functional theory (DFT) calculations reveal the relative stabilities of F substituting on different O sites as a function of the pseudocubic lattice parameter and to elucidate the structural implications of F occupying different sites. Among the three O sites, Figure 1(e) reveals that the relative energy of O1-site substitution steadily decreases with decreasing in-plane lattice parameter compared to O2-site substitution. This reveals that O1-site substitution becomes increasingly favored under compressive strain. The relative orientation of Mn-F bonding with respect to biaxial plane is the prime factor in the site-preference as a function of strain. While the Mn-O2 and Mn-O3 bonds are aligned along in-plane directions, the Mn-O1 bond is the only bond aligned along the *c*-axis (out-of-plane) direction. As compressive biaxial strain decreases the length of the in-plane bonds and elongates the out-of-plane bonds, the longer bond length of Mn-F compared to the Mn-O bond prefers to occur along the out-of-plane direction which stabilizes O1-site substitution. Here, we note that the DFT calculations utilize the oxygen vacancy ordered structure of $Sr_2Mn_2O_5$, while experimentally, we do not have direct evidence of vacancy ordering in the as-grown $SrMnO_{2.5}$ films. However, the physical argument that compressive strain favors creating Mn-F bonds along the growth direction, or oppositely tensile strain favors in-plane Mn-F bonds, holds for the case of ordered or disordered oxygen vacancy distributions. Similarly, the orientation of the films' orthorhombic axes is unknown. Previous studies have shown that $SrMnO_{2.5}$ deposited on LSAT and STO grow with the orthorhombic *c*-axis out-of-the-plane, forming in-plane domains in which the orthorhombic *a*-axis is rotated by 90°.[32] The orientation of the orthorhombic *c*-axis on LAO substrates is unknown, but the preference for F occupancy of apical-like sites under compressive strain would remain even if the orthorhombic *c*-axis of the films lies in-the-plane.

The DFT calculations also reproduce the enhanced *c*-lattice expansion upon fluorination under compressive strain as plotted in Figure 1(f). The O1-site substitution induces an ~20% larger *c*-lattice expansion under fluorination compared to F substitution on the O2 and O3 sites: the *c*-axis expansion with $\gamma = 0.0625$ is 0.07 Å for O1-site substitution and 0.05 Å for O2-site substitution. We note that while the *c*-lattice parameter of the compressively strained SMOF film is largest with F occupying the O1 site, the increased rate of lattice expansion under compressive strain occurs regardless of the substitution site, consistent with a general fluorination-induced lattice volume expansion in which Δ*c* becomes larger with reduced *ab*-plane unit area under compressive strain.

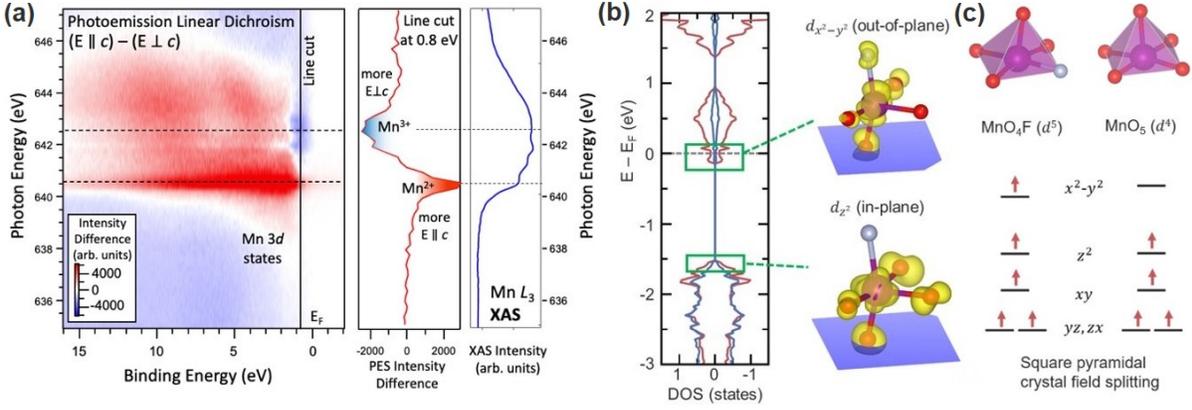

**Figure 2.** Electronic structure measurements and calculations. (a) Resonant polarization-dependent valence-band spectroscopy of SMOF/LAO and the corresponding Mn $L_3$ edge XA spectrum. Vertical line cut at the binding energy of the near-Fermi-level Mn $3d$ states (0.8 eV) illustrates a photon-energy dependent linear dichroism. (b) Density of states of $SrMnO_{2.5-\delta}F_\gamma$ and projected charge density near the Fermi level. The substrate is depicted as the blue plane. (c) Schematic illustration of orbital filling in $MnO_4F$ and $MnO_5$ square pyramidal units.

We used polarization-dependent resonant valence-band photoemission spectroscopy with linear dichroism to study the Mn-derived contributions to the valence band of the SMOF/LAO film. By tuning the photon energy to the Mn $L_3$ absorption edge, we take advantage of the interference between the direct photoemission process $2p^63d^n \rightarrow 2p^63d^{n-1} + e^-$ and the decay of a resonantly excited state $2p^53d^{n+1} \rightarrow 2p^63d^{n-1} + e^-$. This leads to the enhancement of the $3d$ photoionization cross sections, thus effectively amplifying the contribution from the Mn $3d$ states in the valence bands. Linearly polarized x-rays with the electric-field orientation perpendicular ($E \perp c$) and parallel ($E \| c$) to the $c$-axis of the film were used to preferentially probe the in-plane and out-of-plane orbitals. The resonant valence-band linear dichroism map ($E\|c - E\perp c$) is shown in Figure 2(a). By scanning the photon energy across the Mn $L_3$ edge, dichroic features in the near-Fermi-level DOS are observed at 640.5 and 642.4 eV, which are associated with the nominal $Mn^{2+}$ and $Mn^{3+}$ states, respectively. Both features are recognized based on previous resonant photoemission studies on Mn, where $Mn^{2+}$ peak is reported to be at 640.9 eV, and $Mn^{3+}$ is at 643 eV.[33] The incorporation of F into $SrMnO_{2.5-\delta}F_\gamma$ via the topochemical reaction used herein leads to a partial reduction of Mn, with previous x-ray absorption spectroscopy experiments revealing the presence of a mixed $Mn^{2+}/Mn^{3+}$ valence.[19] Thus, the photon-energy dependence of the resonant linear dichroism spectrum shown in the near-$E_F$ line-cut plot indicates that the Mn 3$d$ electrons near the Fermi level prefer to occupy states with an in-plane orbital character at the $Mn^{3+}$ sites and states with out-of-plane orbital character at the $Mn^{2+}$ sites.

DFT calculations reveal that the linear dichroism in the photoemission spectra arises from the orientation of square pyramidal units with respect to substrate. Figure 2(b) shows the density of states (DOS) of SMOF and projected charge density near the Fermi level and valence band edge. The electronic states of Mn near the F atom are well separated, which enables us to identify the orbital character corresponding to $Mn^{2+}$ ($d^5$) and $Mn^{3+}$ ($d^4$) states. The projected charge density clearly shows that $Mn^{2+}$ peak is mostly associated with the $d_{x^2-y^2}$ orbital while $Mn^{3+}$ is associated with the $d_{z^2}$ orbital. As the apical Mn-O bonds in square pyramidal units in $SrMnO_{2.5}$ are aligned along the $ab$-plane, the $d_{z^2}$ and $d_{x^2-y^2}$ orbitals are aligned in-plane and out-of-plane, respectively, following the crystal field splitting for a square pyramidal coordination [Figure 2(c)]. Thus, the electron-doping effect induced by fluorination partially fills the $d_{x^2-y^2}$ orbital aligned perpendicular to the film plane, exhibiting significant polarization dependence of the F-induced $Mn^{2+}$ states. The orientation and the relative energetics of the $d_{z^2}$ and $d_{x^2-y^2}$ orbitals is independent of the F-site occupancy as determined by the crystal field splitting of the square pyramidal coordination, and thus, we attribute the dichroic photoemission response to the combination of the square pyramidal coordination and the presence of some Mn cations in the nominal $d^5$ state.

Linearly-polarized x-ray absorption (XA) spectroscopy carried out at the O and F $K$-edges provide direct evidence for strain-induced changes to the anion site occupation. This approach was recently used to demonstrate anion-site ordering in strained oxynitride $SrTaO_2N$ films.[34] Figure 3 shows the F and O $K$-edge spectra of SMOF on LAO and LSAT measured with $E\|c$ and $E\perp c$ polarized photons. The F contents of these two films are very similar, as shown in Figure 1(d), but they are under opposite strain states. For both SMOF on LAO and LSAT, the F $K$-edge consists of a four-peak spectral shape (which we label $a_1 - a_4$), similar to the previously reported F $K$-edge XA spectra of $MnF_2$ and $MnF_3$,[35] which are shown in Figure S4. The $a_1$ and $a_2$ peaks, located at lower energy positions, are found in the F $K$-edge spectra of $MnF_3$ ($Mn^{3+}$-F), while the $a_3$ and $a_4$ peaks at higher energy positions were previously reported for $MnF_2$ ($Mn^{2+}$-F). The presence of all four absorptive features ($a_1 - a_4$) is consistent with our previous report of the $Mn^{2+}/Mn^{3+}$ mixed valence state in $SrMnO_{2.5-\delta}F_\gamma$ films as revealed through analysis of the Mn $L$-edge spectrum.[19] For the compressively strained SMOF/LAO shown in Figure 3(a), the XA intensities of the $a_1$ and $a_2$ peaks, corresponding to $Mn^{3+}$-F, are significantly enhanced under $E\|c$ polarization. In contrast, the XA intensities of the $a_3$ and $a_4$ peaks, corresponding to $Mn^{2+}$-F, are enhanced under $E\perp c$ polarization. XA spectra of the tensile-strained SMOF/LSAT film shown in Figure 3(b) exhibits the opposite polarization-dependent behavior. The O $K$-edge spectra within the pre-edge energy range of 526 – 534 eV also exhibit opposite trends for the oxyfluorides under the two strain states, shown in Figure 3(d,e).

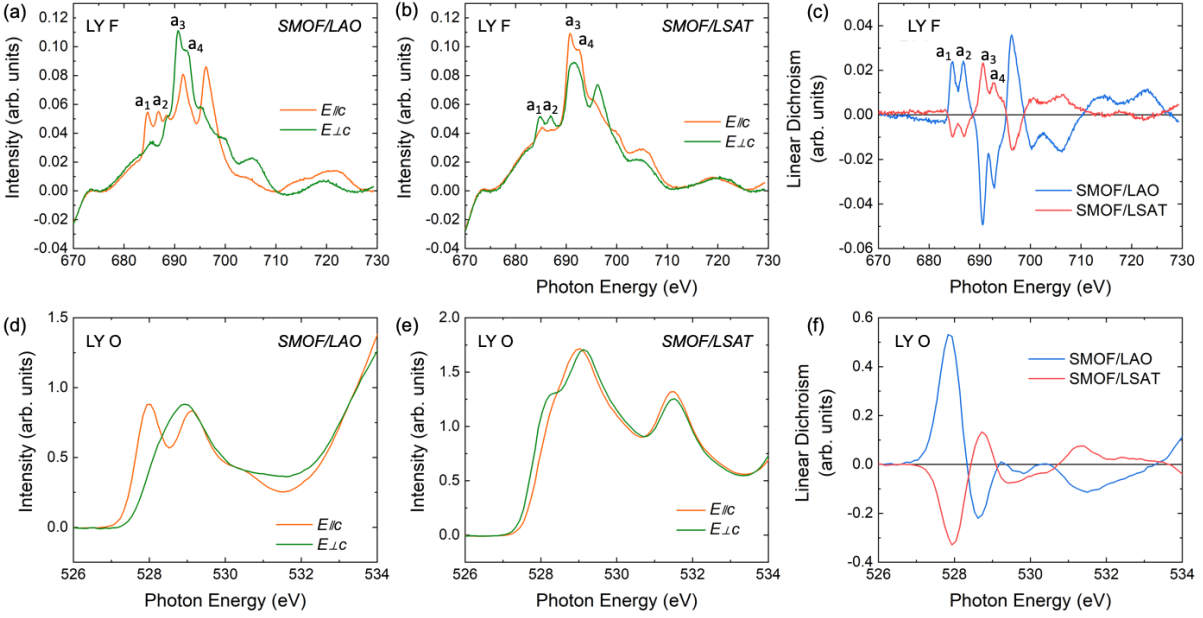

**Figure 3.** X-ray linear dichroism measured on the ligand *K*-edges. The F *K*-edge LY signal of (a) SMOF/LAO, (b) SMOF/LSAT, and (c) F *K*-edge XLD. The O *K*-edge LY signal of (d) SMOF/LAO, (e) SMOF/LSAT, and (f) O *K*-edge XLD. The negative intensities in (a) and (b) originate from the background subtraction procedure.

To directly compare the absorption asymmetry under the two polarization states, we calculate the x-ray linear dichroism (XLD) difference spectra, defined as $E\|c$ - $E\perp c$.[36] The XLD spectra of F *K*-edge for SMOF grown on LAO and LSAT are plotted in Figure 3(c), and the XLD spectra of O *K*-edge for SMOF grown on LAO and LSAT are plotted in Figure 3(f). For both the O and F edges, the XLD of SMOF/LAO and SMOF/LSAT are very similar in shape but opposite in sign. The spectral shape reflects the partial DOS of the unoccupied F 2*p* and Mn 3*d* hybridized orbitals, which are shown schematically in Figure S3 with the orbital energy levels and orientations obtained from DFT calculations. Note that for (001)-oriented SrMnO$_{2.5}$, the $3d_{x^2-y^2}$ and $2p_x$ orbitals point out-of-plane and the $3d_{z^2}$ and $2p_y/2p_z$ orbitals are in-plane as illustrated in Figure 2(b), which is counter to the conventional orbital energetics found in *AB*O$_3$ perovskite films.[37]

We analyze the XLD spectra beginning with the compressively strained SMOF/LAO sample using similar arguments to those recently employed to interpret XLD spectra of strained oxynitride films.[34] For F bonded with $Mn^{3+}$($3d^4$), the lowest energy excitation is to the empty $3d_{x^2-y^2}$ orbital. The electron excitation is allowed to hybridized $2p_x$-$3d_{x^2-y^2}$ and $2p_y$-$3d_{x^2-y^2}$ orbitals contributed by F in O1- and O2/O3-sites, respectively. In the $Mn^{3+}$-F region ($a_1$, $a_2$), the positive XLD of SMOF/LAO indicates larger absorption under the out-of-plane polarization (OPP) condition, where the electric-field vector of the incident x-ray is parallel to the $c$-axis ($E\|c$). This corresponds to an electron excitation into the $2p_x$-$3d_{x^2-y^2}$ orbital, associated with $Mn^{3+}$ bonded to apical-like F. Therefore, F bonded with $Mn^{3+}$ prefers to occupy the O1-sites. For F bonded with $Mn^{2+}$($3d^5$), each orbital is occupied with one electron, and all the electron spins are aligned. When the core electron is excited into the $3d$ states, it has to share the orbital with another electron with opposite spin, which pushes the absorption leading edge towards higher energy.[38] Under the in-plane polarization (IPP) condition, the incident x-rays induce electron excitations into $2p_z$-$d_{zx}$, $2p_z$-$d_{yz}$ and $2p_y$-$d_{yz}$ orbitals, derived from both apical-like (O1-site) and equatorial-like F (O2/O3-sites). Under the OPP condition, the electron excitation is allowed to the $2p_x$-$d_{zx}$ orbital, contributed by equatorial-like F. In the $Mn^{2+}$-F region ($a_3$, $a_4$), SMOF/LAO exhibits a negative XLD, indicating a larger absorption under the IPP condition, which involves contributions of empty states derived from the apical-like F anions. Therefore, there is a higher F occupancy on the O1-sites when bonded with $Mn^{2+}$. Thus, the XLD spectra for compressively strained SMOF/LAO indicates an apical-like site preference for F when bonded to either $Mn^{2+}$ or $Mn^{3+}$.

For the tensile strained SMOF/LSAT, oppositely-signed XLD features are present. In the $Mn^{3+}$-F region ($a_1$, $a_2$), the negative XLD indicates a higher XA intensity under the IPP condition, consistent with electron excitations into the $2p_y$-$3d_{x^2-y^2}$ orbitals derived from equatorial-like F.

In the $Mn^{2+}$-F region ($a_3$, $a_4$), positive XLD indicates higher XA intensity under the OPP condition in which core electron are excited into $2p_x$-$3d_{zx}$ orbitals associated with equatorial-like F. Therefore, F shows a O2/O3-site preference in SMOF/LSAT under tensile strain.

An expected consequence of the observed F site occupancy is that the O site occupancy should exhibit the opposite behavior. For example, an increase of F on the O1-sites should result in a decrease of O on the O1-sites. The O K-edge XLD data shows spectral signatures of this behavior. As shown in Figure 3(f), the O *K*-edge XLD of SMOF grown on LAO and LSAT exhibit an opposite-sign relation similar to the F *K*-edge XLD in Figure 3(c), consistent with strain-induced site preferences for the O ligands.

The full linearly-polarized x-ray absorption spectra for as-grown SMO and fluorinated SMOF on LAO, LSAT, and STO substrates are shown in Figure S5 (TEY collection mode) and Figure S6 (LY collection mode). The TEY signal of F and O XA and XLD has the same spectral shape and positive/negative nature as the LY signal we discussed above. The SMOF/STO film shows similar general XA and XLD spectral features to SMOF/LSAT for both F and O *K*-edges, further confirming the tensile strain induced O2/O3-site preference of F.

From the XLD and DFT results, we conclude that F/O arrangement is controlled by epitaxial strain. This strain-dependence of the anion site-occupancy explains the abnormal *c*-axis lattice expansion of SMOF/LAO even as that sample has the lowest F content. The Mn-F bond length is larger than that of Mn-O,[19] and therefore the O1-site preference of F results in the majority of long Mn-F bonds aligned along the *c*-axis, resulting in a large *c*-axis expansion.

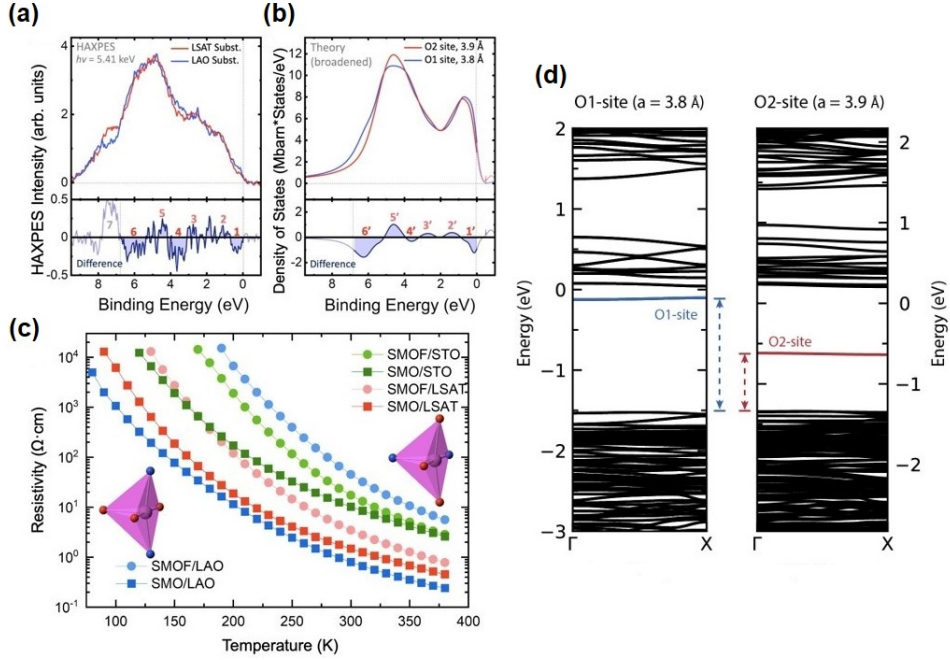

**Figure 4.** Impact of F-site occupancy on electronic properties. (a) Valence band HAXPES spectra measured for SMOF/LAO and SMOF/LSAT thin films. (b) Calculated valence band density of states for SMOF under tensile strain ($a$ = 3.9 Å) with F on the O2-site and under compressive strain ($a$ = 3.8 Å) with F on the O1 site. Difference spectra are also shown with all major features labeled 1–7 for the experiment (a) and corresponding features labeled 1'–6' for the theory (b). (c) Resistivity as a function of temperature of as-grown $SrMnO_{2.5}$ grown on LAO, LSAT, and STO substrates and fluorinated $SrMnO_{2.5-\delta}F_\gamma$ films. (d) Calculated electronic band structures for compressively strained $SrMnO_{2.5-\delta}F_\gamma$ with F on the O1-site and tensile strained $SrMnO_{2.5-\delta}F_\gamma$ with F on the O2-site. The band structures across the full Brillouin zone are presented in the Supplementary Information.

We next turn to the impact of strain-induced anion ordering on the SMOF electronic behavior. The occupied valence band DOS, as measured with hard x-ray photoemission spectroscopy (HAXPES), exhibit differences between the compressive and tensile strained films, as shown in Figure 4(a). To verify that the spectral differences arise from the F-site occupancy, the density of states were calculated using DFT with F residing on the O1-site and a 3.8 Å in-plane lattice parameter and F residing on the O2-site and a 3.9 Å in-plane lattice parameter, mimicking the samples under compressive and tensile strain, respectively [Figure 4(b)]. General agreement is apparent between the measured and calculated spectra down to approximately 7 eV below the Fermi level, as highlighted by the difference spectra between the two strain states presented in the bottom panels of Figure 4(a,b). In particular, spectral features labeled 1-6 all exhibit the same sign at approximately the same energies from the experimental and computational results. Agreement between the HAXPES and DFT results breaks down at binding energies larger than 7 eV, which we attribute to C-based surface contamination on the experimental samples.[39]

The electronic resistivity of as-grown SMO and fluorinated SMOF films on LAO, STO, and LSAT are shown in Figure 4(c); all measurements were performed in a van der Pauw geometry with silver contacts at the sample corners. While all the films are insulators, the change in resistivity upon conversion from SMO to SMOF is dependent on the F-site occupancy. For the as-grown films, the compressive strained SMO/LAO shows the lowest resistivity. The tensile strained films show higher resistivity, with the SMO/STO film being more resistive than SMO/LSAT. After fluorination, all the films become more resistive, this is consistent with our previous report[40] that the resistivity of $SrMnO_{2.5-\delta}F_\gamma$ increases with the increasing amount of F incorporated ($\gamma$). For the tensile strained SMOF films, SMOF/STO shows a larger resistivity increase after fluorination than SMOF/LSAT, which agrees with the previously reported positive correlation between $\gamma$ and resistivity.[40] However, the compressively strained SMOF/LAO film with the lowest F content of the three samples exhibits the largest resistivity and the largest increase of resistivity upon fluorination. This suggests that the F site occupation plays a role in the resistivity of the films. Electronic structure calculations provide insights into a possible origin by which selective anion site substitution impacts the carrier transport. The substitution of F induces a defect state in the band gap; this state is positioned higher in energy when F substitutes on the O1-site compared to the O2-site. This is shown in Figure 4(d) for occupancy of the O1-site under compressive strain and O2-site under tensile strain, although this general trend is observed regardless of applied biaxial strain. The energy difference between the defect state and valence band edge represents the activation energy for electronic conductivity in p-type semiconductors. Here we note that while the energetic magnitude of the two defect states would suggest a larger difference in the resistivity than what we experimentally observe, the qualitative behavior is consistent and there are many additional extrinsic factors (point defect concentrations, orthorhombic domain boundaries, etc) that are not captured in the DFT calculations but could also play a role in the measured resistivity. The carrier effective mass is another factor determining the transport, captured to first approximation by the band curvature. Upon applying compressive strain, we found that the electronic bands of SMO are more dispersive (Figure S7); this smaller effective mass can potentially explain the decreased resistivity of as-grown SMO under compressive strain compared to tensile strain, shown in Figure 4(c). However, fluorinated systems show more localized bands than SMO, and importantly, we found no significant dependence of the band curvature on biaxial strain. Thus, we speculate that the substantially higher resistivity of SMOF/LAO sample results, at least in part, from the energetic differences in the mid-gap levels for F occupying the apical-like anion site (O1-site) compared to the equatorial-like anion site (O2/O3-sites).

## 3. CONCLUSIONS

We have shown that epitaxial strain induces preferential site occupancy for F and O anions in epitaxial $SrMnO_{2.5-\delta}F_\gamma$ oxyfluoride films grown on LAO, LSAT, and STO substrates. X-ray linear dichroism measurements revealed that under compressive strain, F is more likely to take apical-like O1-sites, while under tensile strain, F preferentially occupies the equatorial-like O2/O3-sites.

The site preferences were further confirmed by the calculated relative energies of different F occupancies as a function of biaxial strain. The anionic site-selective substitutions result in changes to the valence band density of states and lead to an increased resistivity in samples in which F occupies the apical-like sites compared to those in which F resides at the equatorial-like position. The control of anion ordering in perovskite oxyfluorides using a readily available substrate-induced strain approach provides a new strategy to tailor crystal symmetries and functional properties in mixed-anion materials.

## 4. METHODS

**Preparation of strained SrMnO$_{2.5}$ films**. The as-grown SMO films were deposited by molecular beam epitaxy on LAO, LSAT, and STO substrates, simultaneously in the same growth to eliminate growth-to-growth differences. The growth temperature was held at approximately 600 °C, while the chamber pressure was maintained at approximately $2.5 \times 10^{-6}$ Torr after introducing O$_2$. Each growth cycle was comprised of a co-deposition of Sr and Mn with a shutter open time of approximately 30 s followed by a 10 s pause, until a targeted thickness of 80 unit cells was achieved. The Sr/Mn stoichiometry was verified by XPS which has been calibrated by Rutherford backscattering spectroscopy.

**Fluorination**. To synthesize oxyfluorides, we use a fluorination method similar to that described by Katayama.[15] The SMO films and fluoropolymers were placed in an alumina boat wrapped by aluminum foil, then the boat was loaded into a horizontal tube furnace with Ar as the carrier gas. Our previous efforts have revealed that using polytetrafluoroethylene (PTFE) as fluorine source produces a uniform F distribution within the SMOF films, and the F content can be controlled by fluorination time and temperature.[19,40] In this study, the strained SrMnO$_{2.5}$ films were fluorinated with PTFE for 4 h at 200 °C. Following fluorination, the samples are kept in air as we have not noticed material degradation in ambient conditions.

**Structural and compositional characterization**. X-ray diffraction (XRD) was performed before and after fluorination using a Rigaku SmartLab diffractometer. The XRD data was simulated in GenX.[41] Reciprocal space maps were obtained from the fluorinated films using beamline 33-BM at the Advanced Photon Source. The atomic concentrations were obtained from XPS sputter depth profiles carried out throughout the entire film thickness. XPS spectra of Sr $3d$, Mn $2p_{3/2}$, O $1s$, and F $1s$ were collected for each measurement cycle after 1 min of $Ar^+$ sputtering using 500 eV ions. After 30 sputtering cycles, the depth profile reached the interface of the film and substrate. In converting from sputtering time to depth, we assume the sputtering rate is uniform with ~ 1 nm of film removed during each sputtering cycle, calculated by dividing the total thickness of the film with the number of sputtering cycles to etch through the entire film. The elemental depth profile was obtained by analyzing the measured photoemission spectra in CasaXPS.[42] The F concentration was determined by averaging the F distribution throughout the film as obtained from XPS depth profiles shown in Figure S2; the presented F concentration ($\gamma$) was obtained by normalizing the F composition to the average of the Sr and Mn composition. We do not have a measure $\delta$ within the films, but the presence of $Mn^{2+}$ after fluorination indicates that $\delta \leq \gamma < 2\delta$.

**X-ray linear dichroism**. The F and O arrangements of the strained SMOF films were evaluated by XLD obtained from linearly polarized XA spectroscopy, performed at Advanced Light Source beamline 4.0.2. The polarized XA spectra of the F $K$-edge, O $K$-edge, and Mn $L$-edge were recorded in near-grazing x-ray incidence geometry with in-plane polarization (IPP, $E\perp c$) and out-of-plane polarization (OPP, $E\|c$) and using both total electron yield (TEY) and luminescence yield (LY) detection modes. The TEY signal is acquired by monitoring the sample drain current originating from the emission of photoelectrons and secondary electrons created by the absorbed x-rays. The inelastic mean-free path (IMFP) of the emitted electrons limits the information depth of the TEY mode to approximately 5 nm, making it a relatively surface-sensitive probe.[43] In contrast, the LY detection mode is bulk sensitive, capturing information from the entire film thickness by monitoring the absorption-modulated visible luminescence of the substrate.[44,45] The XLD spectra were calculated by the difference spectra, defined as XLD = $E\|c$ - $E\perp c$.

**Polarization-dependent resonant valence-band photoemission**. Valence-state dependence of the orbital polarization of the occupied Mn $3d$ orbitals was investigated using resonant valence-band photoemission spectroscopy with linear dichroism at the soft-X-ray ARPES endstation[46] of the high-resolution ADRESS beamline[47] at the Swiss Light Source, Paul Scherrer Institute. All the measurements were carried out with linearly polarized x-rays in the near-grazing incidence geometry, similar to that for XA measurements, thus facilitating selective probing of the in-plane ($E\perp c$) and out-of-plane ($E\|c$) orbitals. The photon energy of the exciting x-ray radiation was scanned across the Mn $L_3$ absorption edge with a step size of 0.1 eV, which is comparable to the total experimental energy resolution (0.11 eV). By tuning the excitation energy to the photon energies of the lower- and higher-lying features at the absorption threshold, the photoionization cross sections of the Mn $3d$ valence-band states occupying the reduced and oxidized Mn cation sites were enhanced, respectively.[48,49]

**Hard x-ray valence-band photoemission**. Bulk-sensitive valence-band spectroscopy measurements were carried out using a laboratory-based HAXPES spectrometer equipped with focused monochromated Cr $K\alpha$ x-ray source and a ScientaOmicron EW4000 electrostatic photoelectron analyzer. At the characteristic excitation energy of 5.4 keV, the IMFP of the photoemitted valence-band electrons in SMOF is estimated to be approximately 7.5 nm,[50] thus facilitating a direct and non-destructive probing of the bulk density of states for a straightforward comparison to theory.

**Density functional theory calculations**. We used the Vienna Ab-initio Simulation Package (VASP)[51,52] to investigate the electronic structure of manganite oxides and oxyfluorides. The calculations utilized the spin-polarized generalized gradient approximation (GGA) with Perdew-Burke-Ernzerhof (PBE) functional, and a Hubbard $U$ correction of 3 eV applied to the Mn 3$d$ orbitals to account for electronic correlations. The projector-augmented wave (PAW) method[53] with a kinetic energy cutoff of 550 eV for the plane wave basis set was used to treat the core and valence electrons using the following configurations Sr ($4s^24p^65s^2$), Mn ($3p^64s^23d^6$), O ($2s^22p^4$), and F ($2s^22p^5$). An 8×4×6 Monkhorst-Pack grid was used for SrMnO$_{2.5}$ with the tetrahedron method while a 4×4×6 grid was used for SrMnO$_{2.5-\gamma}$F$_\gamma$ oxyfluorides. The atomic structures were relaxed until the forces on each atom were less than 5 meVÅ$^{-1}$; the $ab$-plane was subjected to the equivalent pseudocubic lattice parameter to mimic the thin film geometry with biaxial strain.

The atomic structures of the manganite oxyfluorides were constructed from a 2×1×1 supercell of SrMnO$_{2.5}$ by substituting a fluorine atom on one of the different oxygen sites (O1, O2, and O3), which forms SrMnO$_{2.5-\gamma}$F$_\gamma$ with $\gamma$ = 0.0625. The scenario of F occupying vacancy site is not considered for this study because of its low probability as revealed in our previous work.[19] SrMnO$_{2.5}$ adopts a unique zig-zag type spin ordering within the $ab$-plane realized by the square pyramidal units, stabilizing E- and E'-type antiferromagnetic order. The two magnetic orders compete in thermodynamic stability and exhibits similar property trends with varying biaxial strain level, and the detailed comparison of the structural and properties are included in the Supporting Information.

## Supporting Information

The supporting information is available free of charge at [https://pubs.acs.org/…].

X-ray diffraction reciprocal space maps and 00$L$ scans of SMOF films; compositional depth profiles obtained from XPS; schematic of $d$ orbitals and relationship to x-ray polarization; F $K$-edge x-ray absorption of Mn-F compounds; x-ray absorption spectra from Mn $L$-edge, O $K$-edge and F $K$-edge from SMOF films; band structure calculations of $SrMnO_{2.5}$ and $SrMnO_{2.5-\delta}F_\gamma$ as a function of biaxial strain; illustration and relative energies of E and E' magnetic orderings in $SrMnO_{2.5}$ and $SrMnO_{2.5-\gamma}F_\gamma$.

**AUTHOR INFORMATION**

**Corresponding author**: smay@drexel.edu

**Author contributions**

J.W. and S.J.M. conceived the project. J.W. synthesized the samples, performed structural and compositional characterization, measured electronic transport, and assisted with XLD analysis. Y.S. and J.M.R. performed DFT calculations. J.R.P., J.D.G., R.K.S., and A.X.G. performed HAXPES measurements and analysis. J.R.P., J.D.G., R.K.S., A.Z., V.N.S., and A.X.G. carried out polarized resonant photoemission spectroscopy. J.R.P., W.Y., C.K., P.S., and A.X.G. performed XA and XLD measurements. E.K. measured reciprocal space maps via synchrotron x-ray diffraction. J.W., Y.S., A.X.G., J.M.R., and S.J.M. wrote the manuscript with input from the other authors.

**Notes**

The authors declare no competing financial interests.

**ACKNOWLEDGEMENTS**

Work at Drexel was supported by the National Science Foundation (Grant No. CMMI-1562223). Thin film synthesis utilized deposition instrumentation acquired through an Army Research Office DURIP grant (Grant No. W911NF-14-1-0493). Y.S. and J.M.R. were supported by NSF (Grant No. DMR-2011208). Calculations were performed using the Extreme Science and Engineering Discovery Environment (XSEDE), which is supported by National Science Foundation (NSF) Grant No. ACI-1548562, and the Center for Nanoscale Materials, an Office of Science user facility, was supported by the U.S. Department of Energy, Office of Science, Office of Basic Energy Sciences, under Contract No. DE-AC02-06CH11357. J.R.P., R.K.S., and A.X.G. acknowledge support from the U.S. Department of Energy, Office of Science, Office of Basic Energy Sciences, Materials Sciences, and Engineering Division under award number DE-SC0019297. The electrostatic photoelectron analyzer for HAXPES measurements was acquired through an Army Research Office DURIP grant (Grant No. W911NF-18-1-0251). This research used resources of the Advanced Light Source, a U.S. DOE Office of Science User Facility under contract no. DE-AC02-05CH11231.

**REFERENCES**

1  Pena, M.; Fierro, J. Chemical structures and performance of perovskite oxides. *Chem. Rev.* **2001**, *101*, 1981-2018.
2  Schlom, D. G.; Chen, L. Q.; Pan, X.; Schmehl, A.; Zurbuchen, M. A. A thin film approach to engineering functionality into oxides. *J. Am. Ceram. Soc.* **2008**, *91*, 2429-2454.
3  Goodenough, J. B. Electronic and ionic transport properties and other physical aspects of perovskites. *Rep. Prog. Phys.* **2004**, *67*, 1915-1993.
4  Davies, P.; Wu, H.; Borisevich, A.; Molodetsky, I.; Farber, L. Crystal chemistry of complex perovskites: new cation-ordered dielectric oxides. *Annu. Rev. Mater. Res.* **2008**, *38*, 369-401.
5  King, G.; Woodward, P. M. Cation ordering in perovskites. *J. Mater. Chem.* **2010**, *20*, 5785-5796.
6  Rondinelli, J. M.; Fennie, C. J. Octahedral rotation-induced ferroelectricity in cation ordered perovskites. *Adv. Mater.* **2012**, *24*, 1961-1968.
7  Clemens, O.; Slater, P. R. Topochemical modifications of mixed metal oxide compounds by low-temperature fluorination routes. *Rev. Inorg. Chem.* **2013**, *33*, 105-117.
8  Kageyama, H.; Hayashi, K.; Maeda, K.; Attfield, J. P.; Hiroi, Z.; Rondinelli, J. M.; Poeppelmeier, K. Expanding frontiers in materials chemistry and physics with multiple anions. *Nat. Commun.* **2018**, *9*, 772.
9  Kobayashi, Y., Yoshihiro, T.; Kageyama, H. Property engineering in perovskites via modification of anion chemistry. *Annu. Rev. Mater. Res.* **2018**, *48*, 303-326.
10  Harada, J. K., Charles, N., Poeppelmeier, K. R.; Rondinelli, J. M. Materials design: heteroanionic materials by design: progress toward targeted properties. *Adv. Mater.* **2019**, *31*, 1970134.
11  Fuertes, A. Prediction of anion distributions using Pauling's second rule. *Inorg. Chem.* **2006**, *45*, 9640-9642.
12  Schlom, D. G.; Chen, L.-Q.; Fennie, C. J.; Gopalan, V.; Muller, D. A.; Pan, X.; Ramesh, R.; Uecker, R. Elastic strain engineering of ferroic oxides. *MRS Bull.* **2014**, *39*, 118-130.
13  Slater, P. R. Poly(vinylidene fluoride) as a reagant for the synthesis of $K_2NiF_4$-related inorganic oxide fluorides. *J. Fluor. Chem.* **2002**, *117*, 43-45.
14  Moon, E. J.; Xie, Y.; Laird, E. D.; Keavney, D. J.; Li, C. Y.; May, S. J. Fluorination of epitaxial oxides: Synthesis of perovskite oxyfluoride thin films. *J. Am. Chem. Soc.* **2014**, *136*, 2224-2227.

15  Katayama, T.; Chikamatsu, A., Hirose, Y.; Takagi, R.; Kamisaka, H.; Fukumura, T.; Hasegawa, T. Topotactic fluorination of strontium iron oxide thin films using polyvinylidene fluoride. *J. Mater. Chem. C* **2014**, *2*, 5350-5356.

16  Katayama, T., Chikamatsu, A., Kamisaka, H., Kumigashira, H.; Hasegawa, T. Experimental and theoretical investigation of electronic structure of $SrFeO_{3-x}F_x$ epitaxial thin films prepared via topotactic reaction. *Appl. Phys. Express* **2016**, *9*, 025801.

17  Nair, A.; Wollstadt, S.; Witte, R.; Dasgupta, S.; Kehne, P.; Alff, L.; Komissinkiy, P.; Clemens, O. Synthesis and characterization of fluorinated epitaxial films of $BaFeO_2F$: tailoring magnetic anisotropy via a lowering of tetragonal distortion. *RSC Adv*. **2019**, *9*, 37136.

18  Li, J.; Huang, H.; Qiu, P.; Liao, Z.; Zeng, X.; Lu, Y.; Huang, C. Topotactic fluorination induced stable structure and tunable electronic transport in perovskite barium ferrite thin films. *Ceram. Int*. **2020**, *46*, 8761-8765.

19  Wang, J.; Shin, Y.; Arenholz, E.; Lefler, B. M.; Rondinelli, J. M.; May, S. J. Effect of fluoropolymer composition on topochemical synthesis of $SrMnO_{3-\delta}F_\gamma$ oxyfluoride films. *Phys. Rev. Mater.* **2018**, *2*, 073407.

20  Sukkurji, P. A.; Molinari, A.; Reitz, C.; Witte, R.; Kübel, C.; Chakravadhanula, V. S. K.; Kruk, R.; Clemens, O. Anion doping of ferromagnetic thin films of $La_{0.74}Sr_{0.26}MnO_{3-\delta}$ via topochemical fluorination. *Materials* **2018**, *11*, 1204.

21  Kawahara, K.; Chikamatsu, A.; Katayama, T.; Onozuka, T.; Ogawa, D.; Morikawa, K.; Ikenaga, E.; Hirose, Y.; Harayama, I.; Sekiba, D.; Fukumura T.; Hasegawa, T. Topotactic fluorination of perovskite strontium ruthenate thin films using polyvinylidene fluoride. *CrystEngComm*. **2017**, *19*, 313.

22  Chikamatsu, A.; Kurauchi, Y.; Kawahara, K.; Onozuka, T.; Minohara, M.; Kumigashira, H.; Ikenaga, E.; Hasegawa, T. Spectroscopic and theoretical investigation of the electronic states of layered perovskite oxyfluoride $Sr_2RuO_3F_2$ thin films. *Phys. Rev. B* **2018**, *97*, 235101.

23  Chikamatsu, A.; Maruyama, T.; Katayama, T.; Su, Y.; Tsujimoto, Y.; Yamaura, K.; Kitamura, M.; Horiba, K.; Kumigashira, H.; Hasegawa, T. Electronic properties of perovskite strontium chromium oxyfluoride epitaxial thin films fabricated via low-temperature topotactic reaction. *Phys. Rev. Mater*. **2020**, *4*, 025004.

24  Katayama, T.; Chikamatsu, A.; Hirose, Y.; Fukumura, T.; Hasegawa, T. Topotactic reductive fluorination of strontium cobalt oxide epitaxial thin films. *J. Sol-Gel Sci. Technol*. **2015**, *73*, 527-530.

25  Onozuka, T.; Chikamatsu, A.; Katayama, T.; Hirose, Y.; Harayama, I.; Sekiba, D.; Ikenaga, E.; Minohara, M.; Kumigashira, H.; Hasegawa, T. Reversible Changes in Resistance of Perovskite Nickelate $NdNiO_3$ Thin Films Induced by Fluorine Substitution. *ACS Appl. Mater. Interfaces* **2017**, *9*, 10882−10887.

26  Lefler, B. M.; Duchoň, T.; Karapetrov, G.; Wang, J.; Schneider, C. M.; May, S. J. Lithographically-constrained topochemistry of oxide thin films to obtain reconfigurable lateral anionic heterostructures. *Phys. Rev. Mater.* **2019**, *3*, 073802.

27  Caignaert, V., Nguyen, N., Hervieu, M.; Raveau, B. $Sr_2Mn_2O_5$, an oxygen-defect perovskite with Mn (III) in square pyramidal coordination. *Mater. Res. Bull.* **1985**, *20*, 479-484.

28  Caignaert, V., Hervieu, M., Nguyen, N.; Raveau, B. The oxygen defect perovskite $Sr_2Mn_2O_5$: HREM study. *J. Solid State Chem.* **1986**, *62*, 281-289.

29  Berry, F. J.; Bowfield, A. F.; Coomer, F. C.; Jackson, S. D.; Moore, E. A.; Slater, P. R.; Thomas, M. F.; Wright, A. J.; Ren, X. Fluorination of perovskite-related phases of composition $SrFe_{1-x}Sn_xO_{3-\delta}$. *J. Phys.: Condens. Matter* **2009**, *21*, 256001.

30  Aschauer, U.; Pfenninger, R.; Selbach, S. M.; Grande, T.; Spaldin, N. A. Strain-controlled oxygen vacancy formation and ordering in $CaMnO_3$. *Phys. Rev. B* **2013**, *88*, 054111.

31  Chandrasena, R. U.; Yang, W.; Lei, Q.; Delgado-Jaime, M. U.; Wijesekara, K. D.; Golalikhani, M.; Davidson, B. A.; Arenholz, E.; Kobayashi, K.; Kobata, M.; de Groot, F. M. F.; Aschauer, U.; Spaldin, N. A.; Xi, X.; Gray, A. X. Strain-engineering oxygen vacancies in $CaMnO_3$ thin films. *Nano Lett.* **2017**, *17*, 794-799.

32  Kobayashi, S.; Ikuhara, Y.; Yamamoto, T. Labyrinth-type domain structure of heteroepitaxial $SrMnO_{2.5}$ film. *Appl. Phys. Lett.* **2013**, *102*, 231911.

33  de Jong, M. P.; Bergenti, I.; Osikowicz, W.; Friedlein, R.; Dediu, V. A.; Taliani, C.; Salaneck, W. R. Valence electronic states related to $Mn^{2+}$ at $La_{0.7}Sr_{0.3}MnO_3$ surfaces characterized by resonant photoemission. *Phys. Rev. B* **2006**, *73*, 052403.

34  Oka, D.; Hirose, Y.; Matsui, F.; Kamisaka, H.; Oguchi, T.; Maejima, N.; Nishikawa, H.; Muro, T.; Hayashi, K.; Hasegawa, T. Strain engineering for anion arrangement in perovskite oxynitrides. *ACS Nano* **2017**, *11*, 3860-3866.

35  Qiao, R., Chin, T., Harris, S. J., Yan, S.; Yang, W. Spectroscopic fingerprints of valence and spin states in manganese oxides and fluorides. *Curr. Appl. Phys.* **2013**, *13*, 544-548.

36  Huang, D.; Wu, W. B.; Guo, G. Y.; Lin, H.-J.; Hou, T. Y.; Chang, C. F.; Chen, C. T.; Fujimori, A.; Kimura, T.; Huang, H. B.; Tanaka, A.; Jo, T. Orbital ordering in $La_{0.5}Sr_{1.5}MnO_4$ studied by soft x-ray linear dichroism. *Phys. Rev. Lett.* **2004**, *92*, 087202.

37  Tokura, Y.; Nagaosa, N. Orbital physics in transition-metal oxides. *Science* **2000**, *288*, 462-468.

38  Olalde-Velasco, P., Jiménez-Mier, J., Denlinger, J., Hussain, Z.; Yang, W. Direct probe of Mott-Hubbard to charge-transfer insulator transition and electronic structure evolution in transition-metal systems. *Phys. Rev. B* **2011**, *83*, 241102.

39  Himpsel, F., Christmann, K., Heimann, P., Eastman, D.; Feibelman, P. J. Adsorbate band dispersions for C on Ru (0001). *Surf. Sci. Lett.* **1982**, *115*, L159-L164.

40  Wang, J.; Shin, Y.; Gauquelin, N.; Yang, Y.; Lee, C.; Jannis, D.; Verbeeck, J.; Rondinelli, J. M.; May, S. J. Physical properties of epitaxial $SrMnO_{2.5-\delta}F_\gamma$ oxyfluoride films. *J. Phys.: Condens. Matter* **2019**, *31*, 365602.

41  Björck, M.; Andersson, G. GenX: an extensible x-ray reflectivity refinement program utilizing differential evolution. *J. Appl. Crystallogr.* **2007**, *40*, 1174-1178.

42  Walton, J., Wincott, P., Fairley, N.; Carrick, A. Peak Fitting with CasaXPS: A Casa Pocket Book. (Acolyte Science, Cheshire, 2010).

43  Stöhr, J. *NEXAFS spectroscopy*. Vol. 25 (Springer Science & Business Media, 2013).

44  Bianconi, A., Jackson, D.; Monahan, K. Intrinsic luminescence excitation spectrum and extended x-ray absorption fine structure above the K edge in $CaF_2$. *Phys. Rev. B* **1978**, *17*, 2021.

45  Li, B.; Chopdekar, R. V.; N'Diaye, A. T.; Mehta, A.; Byers, J. P.; Browning, N. D.; Arenholz, E.; Takamura, Y. Tuning interfacial exchange interactions via electronic reconstruction in transition-metal oxide heterostructures. *Appl. Phys. Lett.* **2016**, *109*, 152401.

46  Strocov, V.; Wang, X.; Shi, M.; Kobayashi, M.; Krempasky, J.; Hess, C.; Schmitt, T.; Patthey, L. Soft-x-ray ARPES facility at the ADRESS beamline of the SLS: Concepts, technical realisation and scientific applications. *J. Synchrotron Radiat.* **2014**, *21*, 32-44.
47  Strocov, V. N.; Schmitt, T.; Flechsig, U.; Schmidt, T.; Imhof, A.; Chen, Q.; Raabe, J.; Betemps, R.; Zimoch, D.; Krempasky, J.; Wang, X.; Grioni, M.; Piazzalunga, A.; Patthey, L. High-resolution soft X-ray beamline ADRESS at the Swiss Light Source for resonant inelastic X-ray scattering and angle-resolved photoelectron spectroscopies. *J. Synchrotron Radiat.* **2010**, *17*, 631-643.
48  Park, J.-H.; Chen, C. T.; Cheong, S.-W.; Bao, W.; Meigs, G.; Chakarian, V.; Idzerda, Y. U. Electronic aspects of the ferromagnetic transition in manganese perovskites. *Phys. Rev. Lett.* **1996**, *76*, 4215.
49  Yamamoto, K.; Horiba, K.; Taguchi, M.; Matsunami, M.; Kamakura, N.; Chainani, A.; Takata, Y.; Mimura, K.; Shiga, M.; Wada, H.; Senba, Y.; Ohashi, H.; Shin, S. Temperature-dependent Eu $3d-4f$ x-ray absorption and resonant photoemission study of the valence transition in $EuNi_2(Si_{0.2}Ge_{0.8})_2$. *Phys. Rev. B* **2005**, *72*, 161101.
50  Tanuma, S., Powell, C.; Penn, D. Calculations of electron inelastic mean free paths. IX. Data for 41 elemental solids over the 50 eV to 30 keV range. *Surf. Interface Anal.* **2011**, *43*, 689-713.
51  Kresse, G.; Furthmüller, J. Efficiency of ab-initio total energy calculations for metals and semiconductors using a plane-wave basis set. *Comput. Mater. Sci.* **1996**, *6*, 15-50.
52  Kresse, G.; Furthmüller, J. Efficient iterative schemes for ab initio total-energy calculations using a plane-wave basis set. *Phys. Rev. B* **1996**, *54*, 11169-11186.
53  Blöchl, P. E. Projector augmented-wave method. *Phys. Rev. B* **1994**, *50*, 17953-17979.